\documentclass[graybox]{svmult}

\usepackage{mathptmx}       % selects Times Roman as basic font
\usepackage{helvet}         % selects Helvetica as sans-serif font
\usepackage{courier}        % selects Courier as typewriter font
\usepackage{type1cm}        % activate if the above 3 fonts are
\usepackage{makeidx}         % allows index generation
\usepackage{graphicx}        % standard LaTeX graphics tool
\usepackage{multicol}        % used for the two-column index
\usepackage[bottom]{footmisc}% places footnotes at page bottom

\usepackage{natbib}
\usepackage{hyperref}

\makeindex             % used for the subject index
\begin{document}

\title*{Deriving High-Precision Radial Velocities}
% Use \titlerunning{Short Title} for an abbreviated version of
% your contribution title if the original one is too long
\author{Pedro Figueira}
% Use \authorrunning{Short Title} for an abbreviated version of
% your contribution title if the original one is too long
\institute{Pedro Figueira \at Instituto de Astrof\'isica e Ci\^encias do Espa\c{c}o, Universidade do Porto, CAUP, Rua das Estrelas, 4150-762 Porto, Portugal,\\ 
\email{pedro.figueira@astro.up.pt}}
%\and Name of Second Author \at Name, Address of Institute \email{name@email.address}}
%
% Use the package "url.sty" to avoid
% problems with special characters
% used in your e-mail or web address
%
\maketitle

\abstract{This chapter describes briefly the key aspects behind the derivation of precise radial velocities. I start by defining radial velocity precision in the context of astrophysics in general and exoplanet searches in particular. Next I discuss the different basic elements that constitute a spectrograph, and how these elements and overall technical choices impact on the derived radial velocity precision. Then I go on to discuss the different wavelength calibration and radial velocity calculation techniques, and how these are intimately related to the spectrograph's properties. I conclude by presenting some interesting examples of planets detected through radial velocity, and some of the new-generation instruments that will push the precision limit further.}

\section{Precise radial velocities}\label{Sec:Intro}

Radial velocities are, by definition, the velocities measured along a given line of sight, and often refer to a velocity calculated through the measurement of the Doppler shift of a given spectral line. In its non-relativistic form, the well-known Doppler shift formula,
\begin{equation}
 \frac{\Delta \lambda}{\lambda} = \frac{v}{c} \, ,
\end{equation}
relates the displacement $\Delta \lambda$ of a line of wavelength $\lambda$ to a radial velocity $v$, with $c$ being the speed of light in vacuum. Measuring a radial velocity (henceforth RV) is fundamentally different from measuring directly a velocity on the plane of the sky, in the sense that the error on the RV does not depend geometrically on the distance to the source as it does for physical velocities measured on the plane of the sky. Instead, the error on the RV depends only on the noise present in the spectrum, and how this noise translates to an uncertainty on the line shift value.

At this point, it is extremely important to clarify what we mean by \textit{precision}, and to characterize the associated type of error. The precision of a measurement system, also called reproducibility or repeatability, is the degree to which repeated measurements performed under unchanged conditions lead to the same results. This repeatability error can be associated to a value measured by the system, becoming the precision of the value. This is conceptually different from the accuracy of a measurement system or value, which is how close a measurement of a quantity is to its real --- true --- value. In several astronomical studies, one is concerned about accurate RVs; here we are only concerned with precise RVs.

Precise RVs have been used to calculate the velocities of stars and study, for instance, Galactic kinematics, stellar binarity, and determine stellar masses. For these scientific objectives, an overall precisison of the order of the km/s was enough. More recently, the presence and characterization of exoplanets was possible when the precision threshold crossed the level of 50--100 m/s. On the other hand, to measure stellar oscillations, differential line shifts, and line profile variations, the asteroseismic studies routinely require a precision on the order of 1--100 m/s, often measured in individual lines as opposed to the whole spectrum, as has been done for exoplanetary searches or binary characterization. These examples serve to illustrate how different scientific objectives require different precision level. But to understand how to derive precise RVs, we will have to understand how a spectrograph works.

\section{Breaking down a spectrograph}\label{Sec:Spectro}

A spectrograph is a scientfic instrument that receives the light collected by a telescope, disperses it, forming a spectrum, and records this spectrum on a detector. But what is inside it? In a conceptual way, a spectrograph is composed of four types of components:
\begin{itemize}
 \item The light interface/feeding with the telescope --- a slit or a fiber;
 \item The dispersive elements (main and secondary) --- like prisms, grisms;
 \item The detector (usually a CCD or CMOS) and its camera;
 \item The optics.
\end{itemize}

The light interface or light feed of the spectrograph has a double function. The first one is to select the target (or targets) of interest in the field of view of the telescope, so that only the selected target's light is fed into the spectrograph. The second one is to define spatially the spectrograph's resolution element: it is the image fed to the spectrograph that will be dispersed as a function of wavelength, and ultimately projected onto the detector. The first element of the spectrograph defines this first (crucial) image. The light interface can be a \textit{slit} or a \textit{fiber}.

A slit is a mechanical aperture with two parallel jaws, allowing one to select a rectangular image from the field of view of the telescope to feed into the spectrograph. This rectangular image is often longer than wider, and its dimensions can be adjusted mechanically; for instance, often one of the slit jaws is fixed and the position of the second one is adjustable, allowing one to change the slit width on the fly. The rectangular image created by the slit will be dispersed along the direction of the slit width --- named \textit{dispersion direction}. Importantly, this setup preserves one direction of the image, the direction perpendicular to the dispersion --- named \textit{spatial direction}. This opens an interesting set of possibilities, allowing one to position the slit in creative ways to feed several different objects and thus record simultaneously different spectra with the spectrograph. However, it is important to remember that the same target (or ensemble of targets), when positioned in different ways on the slit, will lead to different light distributions on the rectangular image of the slit. These different images will then by dispersed by the spectrograph, leading to different spectra as recorded on the detector. As such, it is important to bear in mind that different illumination and light collection patterns of the same targets on the slit will lead to different recorded spectra.

The fiber addresses this illumination aspect directly. A fiber is simply a wave-guide that works based on the total internal reflection principle \citep[see, e.g.,][]{2008SPIE.7018E..4WA, 2010SPIE.7739E..47C}. It is usually composed of a fused silica core and a protective cladding. One end of the fiber is placed at the image created on the focal plane of the telescope. The other end of the fiber will feed the light to the spectrograph. In this setup, it is the image formed at the exit of the fiber that will be dispersed by the spectrograph. As the light rays undergo several reflections inside the fiber, the light distribution is scrambled, losing the memory of its initial distribution (for a circular fiber this scrambling is more efficient in the azimuthal direction, being rather imperfect in the radial one). This scrambling is quantified by the \textit{scrambling gain}, one of the main properties of the fibers, the others being their spectral transmission window and efficiency, the attenuation as a function of fiber length and the focal ratio degradation. Other than the reduction of the impact of illumination variation on the recorded spectra, by using a fiber one can move the spectrograph away from the telescope focus. This allows one to develop heavier, larger, and more stable spectrographs, as done in recent years. Also, by using different fibers one can inject light from several sources onto the same spectrograph simultaneously, exploring interesting concepts like that of UVES+Flames \citep[e.g.,][]{2002Msngr.110....1P}, or simply allowing to use a calibration cell at the same time as we observe our target star. This said, the main disadvantage of using a fiber follows from its ability to scramble light efficiently: all the light sources inside the field of view of the fiber are scrambled and fed to the spectrograph simultaneously, creating a composite spectrum. One loses the ability to identify the spectra associated to each of the targets. Moreover, when adding an extra optical element, there is a fraction of light that is lost at the fiber interface due to reflection from it. This is minimized by using anti-reflection coatings, but a small loss is always present.

The second type of components that exist in all spectrographs are, very naturally, the dispersive elements. These dispersive elements can be \textit{prisms} or \textit{gratings}. A prism is the simplest dispersive element one can conceive. It is a refracting optical element that, through Snell's law and the fact that the refraction index \textit{n}\,=\,\textit{n}($\lambda$), disperses the light that strikes one of its faces. For a detailed description on how the angle of dispersion depends on the wavelength $\theta$($\lambda$), the reader is referred to, e.g., \cite{1987sdap.book.....S}. Unfortunately, the angle of dispersion that can be achieved with such optical devices is very low, and as a consequence the ability to disperse light into different wavelengths is rather limited. As such, one has to resort to more complex optical devices to create spectrographs with the ability to disperse light to resolve fine-scale spectra. 

The most efficient dispersive element is arguably the diffraction grating. As written in the \textit{Newport Diffraction Grating Handbook}\footnote{Which can be found, e.g., at \url{http://optics.hanyang.ac.kr/~shsong/Grating\%20handbook.pdf}.} (one of the most renowned gratings manufacturers): 
\begin{quotation}
A diffraction grating is a collection of reflecting (or transmitting) elements separated by a distance comparable to the wavelength of light under study. It may be thought of as a collection of diffracting elements, such as a pattern of transparent slits (\ldots) or reflecting grooves (\ldots).
\end{quotation}
This device allows for much larger separation angles than prisms, and from the diffraction equation for gratings we have that  
\begin{equation}\label{Eq:grating}
m \lambda = d (\sin{\alpha} + \sin{\beta}) \, ,
\end{equation}
from which it follows 
\begin{equation}
\beta(\lambda) = \arcsin{\frac{m \lambda}{d} - \sin{\alpha}} \, ,
\end{equation}
%NOTE TO EDITORS. If I use the environment ``eqnarray'' the labeling is lost
%\begin{eqnarray}
%m \lambda = d (\sin{\alpha} + \sin{\beta}) \Leftrightarrow \\
%\beta(\lambda) = \arcsin{\frac{m \lambda}{d} - sin{\alpha}}
%\end{eqnarray}\label{Eq:grating}
where $\alpha$ and $\beta$ are the incident and diffracted rays' angles as measured relative to the vertical of the grating, $\lambda$ the wavelength of the light, $m$ the order of interference, and $d$ the separation between groves. The angular dispersion can be found by differentiating the last equation:
\begin{equation}
\label{Eq:grating2}
 \frac{d\beta}{d\lambda} = \frac{m}{d\cos{\beta}} = \frac{\sin{\beta} + \sin{\alpha}}{\lambda \cos{\beta}} \, .
\end{equation}

Equation (\ref{Eq:grating2}) shows that for a given $\lambda$ the angular dispersion depends only on $\alpha$ and on $\beta$. It is clear, however, that Eq.~(\ref{Eq:grating}) and those derived from it have multiple solutions. One is of particular interest for us, called \textit{Littrow}, for which $\alpha$\,=\,$\beta$, i.e., the light is dispersed along the same direction as that of incoming rays. However, the existence of solutions with different $m$ leads to a superposition of different orders along a given $\beta$. One can get around this issue and select the order of interest (for instance, through the usage of filters, or by making the grating much more efficient for a given $m$ than for others), but a more efficient and elegant solution can be achieved by using a second dispersive element. One can \textit{cross-disperse} the orders, dispersing the overlapping orders in a direction perpendicular to the first dispersion (done by the grating). The first disperson is called \textit{main dispersion} and the second the \textit{secondary dispersion}. The secondary dispersion has the objective of separating physically the already dispersed orders, and as such a less powerful dispersive element can be used. By applying the two dispersions and focusing the image of the dispersed orders on the spectrograph, one obtains a ``ladder-like'' pattern, in which the orders are (approximately) parallel to each other. The wavelengths increase along each order, but also from order to order. This pattern gave the name to one of the most used types of high-dispersion gratings in the market, the \textit{echelle grating}. 

For a grating for which the grooves are perfectly aligned on a plane, and applying some simple ray-tracing geometry to Eq.~(\ref{Eq:grating}) and the configuration behind it, one gets that the maximum efficiency of the grating occurs for $m$\,=\,0 (i.e., reflection), decreasing fast as $\left|m\right|$ increases (i.e., the interference with high orders). Since we are often interested in working in \textit{Littrow} or \textit{quasi-Littrow} condition (i.e., the refracted angle is only slightly different for the incident angle, for practical purposes), this forces us to work at high $m$. To avoid working in a very low-efficiency regime of the grating, one can change the geometry of the grooves, by adjusting their angle relative to the grating surface. To make the \textit{Littrow} condition angle the angle with the highest transmission, we can introduce a so-called \textit{Blaze} angle $\delta$ between the grooves and the grating surface. The efficiency is maximized for the \textit{Littrow} angle when $\alpha$\,=\,$\beta$\,=\,$\delta$, and this is called the \textit{Littrow Blaze condition}.

A third and very important part of the spectrograph --- and any astronomical instrument, for that matter --- is the detector. The detector is very simply a device that transforms the incident light into electric charge, usually by photoelectric effect. They are often 2-dimensional, being plane or approximately so, and thus allowing one to record the 2-dimensional image focused by the convergence of the dispersed light. The detectors work by photoelectric effect, through which the arrival of a photon at a given pixel will lead to the production of an electric charge. The measurement of the electric charge as a function of position allows one to map the incident photons, and create an electronic image. The main property to take into account is then, very naturally, the photosensitive material. Different elements and mixtures of elements will have different valence gaps and valence energies, i.e., they will have minimum energies by which they are sensitive to photoelectric effect. For a long while Silicon-based architectures dominated the market of detectors. The ability to create homogeneous grids of photosensitive Silicon (often with Boron), along with the ability to store and transfer the charge generated, opened the way to the \textit{Charge-Coupled Device} (CCD) architecture. In this architecture the charge generated in the pixels is transferred to a common amplifier and register, where it is amplified and read. The usage of a single reading port greatly homogeneizes the detector, allowing the characterization with a single gain value (ability to transform electric charge in readable digital units), readout noise (error introduced by the reading process), along with other key properties. However, the Silicon atom is only sensitive to photons with wavelengths shorter than $1.1\,{\rm \mu m}$ (and detectors made of it are often of very low efficiency for $\lambda>1\:{\rm \mu m}$). To observe at longer wavelengths one must then go for different photosensitive materials. The first and obvious drawback is that these detectors have to be operated at a much lower temperature, for they are more sensitive to their own thermal radiation, and low-energy photons and electrons in general. The second much less obvious feature is that there is no equivalent of Silicon for these longer wavelengths, i.e., there is no material that can be manipulated electronically to transfer the charge to a common amplifier and reader. This means that near-infrared detectors have to perform the charge amplification and reading in-pixel. This leads to much more complex electronics, and the architecture behind it is called \textit{Composite Metal Oxide Semiconductors} (CMOS).  While significantly more complex, the fact that the charge is manipulated in-pixel provides very interesting options. For instance, the charge can be read multiple times to reduce readout noise, or one can read only specific parts of the detector, two options that are unavailable when reading implies a clocked transfer of charge, as for CCDs. The price for the local conversion of charge into voltage is a lower degree of homgeneity in the gain and error, and the more complex electronics often conduce to higher readout noise and spurious currents. However, the increasing demand of CMOS led to a fast development of the technology associated to it, and in several situations CMOS, when available, are already being preferred over the CCD technology. A common example is that of acquisition or guiding cameras, in which the readout time is a critical aspect; one can easily accept a noisy detector if it can be read much faster (through customized window reading, for instance) than its less noisy counterpart.

The last important part of a spectrograph is its optics. The optics accomplish different functions:
\begin{itemize}
 \item to transform the convergent rays of light, focused at the entrance of the spectrograph by the telescope optics, into collimated light;
 \item to transform the dispersed (but still collimated) light into focused light that can be recorded in the detector;
 \item to create a spectral format such that all the orders and wavelengths can be recorded on the detector.
\end{itemize}
These correspond, basically, to the collimation of the light after it enters the spectrograph, and the transformation of the dispersed light into a focused image on the camera. These tasks have to be performed in such a way that the recording of the light is practical and eases data analysis, while maintaining the most desirable properties of the spectrograph and spectra. And this is exactly the point we will address next.

\section{Spectrograph's properties}

The spectrograph's design will define its properties, which in turn will be translated directly to the spectra it forms. The most important spectrograph properties, that can be considered as properties of the associated spectra, are:
\begin{itemize}
 \item Wavelength range;
 \item Transmission (or efficiency);
 \item Resolution;
 \item Sampling;
 \item Instrumental Profile Characteristics (especially shape and stability). 
\end{itemize}

The first two properties are rather self-explanatory. The wavelength range is the wavelength domain in which the spectrograph operates, and the domain of the formed spectra. The choice of wavelength has very important consequences not only on the detectors, as seen before, but on all optical and dispersive elements (requiring different coatings to reduce unwanted reflections and scattered light, for instance). Associated to this first property is the transmission of the spectrograph or its efficiency, which can be defined in terms of total energy or photon number. The transmission is the fraction of energy or photons that traverses the spectrograph and is recorded by the detector\footnote{Importantly, spectrographs with variable slit width separate the transmission of the spectrograph into \textit{transmission of the spectrograph} $\times$ \textit{transmission of the slit}, and detail the transmission of the latter as a function of its (tunable) properties.}. 

The property that follows is a key one: the resolution. The resolution, defined as $R \equiv \Delta \lambda / \lambda$, relies on the \textit{Rayleigh criterion} to establish $\Delta \lambda$ for a given $\lambda$: it is the smallest difference in wavelength between two lines of equal intensity that can be discernible using the spectrograph, i.e., it is the smallest difference in wavelength for which two lines of equal intensity are resolved. Fortunately, there is a different way of understanding the resolution of a spectrograph. A spectrograph can be thought of as a device that convolves an infinite-resolution spectrum, coming from the source, with the instrumental profile (IP) that characterizes the spectrograph. The width $\Delta \lambda$ of this instrumental profile, as measured at a given $\lambda$, defines its resolution $R$. This means that, in practice, to measure the resolution of a spectrograph we can use a line with a full width at half maximum $FWHM$ such that $FWHM$/$\lambda$ $\ll$ 1/$R$. In other words, we are feeding into the spectrograph a line with a width much smaller than the spectrograph's IP. As a consequence the width of the line that results from the convolution, is defined by the spectrograph's IP only, and its $FWHM$ defines the resolution. It is never too much to stress the impact of the resolution on the final spectra, and the reader is invited to look at several examples by him/herself. 

A point which is often overlooked is that of \textit{sampling} (also referred to as \textit{numerical resolution} or \textit{numerical sampling}, or even more obscure names). Sampling is the number of pixels used to record the wavelength interval covering one resolution element of the spectrograph. An application of the Nyquist theorem to this situation informs us that the sampling should be larger than 2 pixels in order to avoid losing a significant fraction of the information. Most modern spectrographs opt for a sampling value of 3 or similar.

The final point in our list is the most intricate one. The instrumental profile (IP) is the profile that represents the broadening introduced by the spectrograph relative to a conceptual infinite-resolution spectrum emitted by the source. By construction, all elements that constitute the spectrograph have an impact on the IP. An ideal spectrograph should introduce a broadening that depends on its resolution, but no other deformation to the spectrum; from that it follows that the IP should be as symmetric as possible, maintaining the original profile of the lines. It is often conceptualized as a positive-definite Gaussian function of unit area. The IP shape should also be independent of wavelength, a condition which is not respected to some degree due to the presence of optical aberrations. Finally, and very importantly, this IP should be as stable as possible, being virtually independent of time. Any variation on the IP will be imprinted on the observed lines, and as such will have an impact on the measurement of the characteristics of the lines, like the RV. This IP stability is arguably the most difficult condition to characterize and ensure, and has a significant impact on the RV, as we will soon see. 

\section{Radial-velocity precision}\label{Sec:PrecisionRV}

We finally arrive at the issue of RV precision. What are the instrumental factors and stellar parameters that have an impact on it? And how can we design both a spectrograph and observations for the best achievable RV? These are the two questions that we ask ourselves in this chapter.

\subsection{Precision achievable on a given spectrum}

The first and most fundamental question to ask is what is the ultimate precision one can achieve when calculating the RV on a given spectrum. This ``floor level'' of precision is the value one achieves when one considers as the only source of error the noise present in the spectrum; for high-$SNR$ spectra, this is the stellar photon noise. The spectrograph measuring the RV is considered as perfect, and the act of measurement introduces no noise. 

Based on the work of \cite{1985Ap&SS.110..211C}, and assuming that a spectrum experiences a differential shift $\delta \lambda$ relative to its own noise-free reference copy, \cite{2001A&A...374..733B} calculated that the optimal weight $W(i)$ to be given to a pixel $i$ when calculating the RV is
\begin{equation}\label{Eq:RVweight}
 W(i) = \frac{\lambda^2(i) (\delta A_0(i) / \delta \lambda)^2}{A_0(i) + \sigma^2_D} \, ,
\end{equation}
where the spectrum (perceived simply as a group of consecutive pixels) is represented by the function ($x(i)$, $y(i)$) = ($\lambda(i)$, $A_0(i)$), and $\sigma_D$ represents the readout error\footnote{This error can be represented as a function of pixel $i$, becoming $\sigma_i$ and even characterize other sources of error, without loss of generality.}. It is important to notice that the denominator of Eq.~(\ref{Eq:RVweight}) is the variance of the flux $A_0$ at a given pixel, when considering both photon noise and readout noise. The error (or scatter) on the measured RV, $\delta v$, can then be calculated as
\begin{equation}
 \delta v = \frac{c}{\sqrt{\sum_i W(i)}} \, .
\end{equation}

Arguably, the most striking aspect about these equations is that the weight of a pixel, for a given $\lambda$, depends on the absolute value of the slope of the spectrum ($\delta A / \delta \lambda$). A larger slope value leads to a larger weight, or to put it differently, the RV information content in a spectrum is contained in the slope of its lines. Our ability to measure the position of a spectrum relative to its own copy depends on the slope of its lines. The sharper the lines, the higher our precision. 

However, this is just one way of measuring the ultimate precision achievable on a given spectrum. \cite{1992ESOC...40..275H} took a rather different approach: assuming that the noise on the spectrum is photon noise only, and that the spectrum is characterized by an uniform density of lines, the authors concluded that the RV error $\sigma_{\rm RV}$ is given by
\begin{equation}\label{Eq:HatzesCochran}
 \sigma_{\rm RV} \propto \frac{1}{\sqrt{F} \; \sqrt{\Delta \lambda} \; R^{1.5}} \, ,
\end{equation}
where $F$ is the average flux level, $\Delta \lambda$ is the wavelength coverage and $R$ the resolution. The term $\sqrt{F}$ represents the photon noise error calculated from the flux, and the $\sqrt{\Delta \lambda}$ represents the increase in statistics represented by including independent measurements of lines. The error on the average RV calculated from $N$ different lines can be thought of $\overline{\sigma} \propto \sqrt{N}$ and as such $\propto \sqrt{\Delta \lambda}$. The new insight from this formula comes from the term $R^{1.5}$: the RV precision depends more steeply on the resolution of the spectrograph than on any of the other mentioned factors. This brings us back to the concept of RV information content and how it is contained in the slope of the lines. When one increases the resolution, the slope of the spectral lines is increased due to an increase in both the line's contrast and a reduction of the line's width.

These dependencies on key parameters are very informative, but it is also interesting to understand how precision changes for a given spectral line, and the simplest assumption one can make about a spectral line is to approximate it by a Gaussian function. Assuming a Gaussian-shaped line and applying the formalism of \cite{2001A&A...374..733B}, one can calculate that the RV precision is given by
\begin{equation}
 \sigma_{\rm RV} = \frac{(\pi\,\ln{2})^{-1/4}}{2} \frac{\sqrt{FWHM}}{SNR} \frac{\sqrt{PXLSC}}{C}F(C_{\rm eff}) \left[{\rm m\,s^ {-1}}\right] \, ,
\end{equation}
where $C$ is the contrast of the Gaussian line, $SNR$ the signal-to-noise ratio of the spectrum at hand, and $PXLSC$ the pixel scale of the spectrograph (i.e., the dimension of the pixel as measured in velocity). $F(C_{\rm eff})$ is a polynomial function of the effective contrast $C_{\rm eff}\,=\,C/(1+\sigma_D^2/A_0)$. This equation shows us very important basic properties, namely:
\begin{itemize}
 \item The RV precision increases linearly with the $SNR$ with which we measure the spectrum;
 \item The RV precision is proportional to the contrast $C$ of a line, and inversely proportional to the square root of the $FWHM$ of the line.
\end{itemize}
These two aspects were already represented in Eq.~(\ref{Eq:HatzesCochran}). Yet, now the $SNR$ dependence is written explicitly (and one can consider noise contributions other than photon noise), and the impact of resolution is broken down into the two characteristics of the lines that it changes: $FWHM$ and $C$. And of course these two characteristics are exactly the key parameters that regulate the slope present in a line. We have made full circle, coming back to the first conclusion brought by \cite{2001A&A...374..733B}.

We have been looking at the impact of the spectrograph's resolution on the achievable RV. As said before, the line shape associated to observing with a given resolution can be seen as the result of the convolution of the stellar spectrum with the IP. This said, another effect has exactly the same impact on the line shape, and as such on the RV: stellar rotation. Stellar rotation, and the line broadening associated to it, is often modeled through the convolution of (non-rotating) stellar spectra with a rotational kernel, a function depending solely on the projected rotational velocity, $v\,\sin{i}$, of the star. As such, it comes as no surprise that $\sigma_{\rm RV} \propto (v\,\sin{i})^{1.5}$, when other line broadening mechanisms are negligible. And this is the reason why on fast rotators we always obtain very poor RV precision. 

\subsection{Spectrographs for precise radial velocities}

We have seen how the properties of a spectrograph define the properties of the spectra acquired with it. Of all these properties, we discussed at length the non-trivial impact of the resolution on the recorded line shape, and in turn the impact of the final line shape on the achievable RV. We saw how the impact of resolution results or can be understood as the convolution with an IP that represents a spectrograph. We also stressed a key point: \textit{Any change in the IP will lead to a change in line shape, which can translate into a measured RV variation.} The realization of this key aspect of the spectrograph's operation led to two schools of thought on how to handle IP-induced RV:
\begin{enumerate}
 \item Control the IP as much as possible, reducing to a minimum its variations as a function of time so that one can reduce its impact on the RV;
 \item Allow the IP to vary but model its variation and remove its effect on the measured spectrum.
\end{enumerate}
The two approaches lead to completely different technical choices, and we discuss each one of them in detail. 

\subsubsection{IP control}

When trying to control the IP variation, one has to act accross the whole spectrograph. As seen in Sect.~\ref{Sec:Spectro}, the first important aspect is that of light injection. Instruments aiming at controlling their IP should use light-scrambling devices, such as fibers, to reduce the spatial effects of variable illumination on RV precision. These variable illumination effects can come from imperfect centering, guiding problems, or simply variable seeing. 

On top of the special care taken with illumination and light-feeding aspects, the whole instrument is designed to ensure its IP is as stable as a function of time as possible. This translates into building instruments that operate under vacuum, and are pressure- and temperature-controlled. For reference, HARPS is stabilized in pressure and temperature at 0.01\,mbar and 0.01\,K, respectively. 

While the whole instrument is carefully monitored to reduce the IP changes, very subtle profile variations can occur, especially over long timescales, over which the physical parameters control can exert a smaller leverage. To monitor these comparatively small IP variations, the instruments are often built so that one can record the spectrum of a calibration source simultaneously with the scientific target. This simultaneous calibration is obtained through two sets of orders, recorded interweaved on the detector. One of the two sets is fed by light collected by the telescope on a science target, and the other set comes from a calibration lamp or device located in a calibration unit. The reference spectrum coming from this second fiber allows one to define on real time a wavelength calibration on the detector, and evaluate how this calibration changes with time. If one assumes that the two sets of orders (or fibers, as one prefers) experience the same IP changes, one can evaluate the wavelength calibration changes in the reference and apply them to the scientific channel. In its simplest form, this corresponds to using the reference to measure an RV drift that can be applyed to each exposure to correct the wavelength calibration relative to wavelenength solution obtained at the beginning of the night, on the first fiber.

If we have the spectrum recorded in our spectrograph with a wavelength calibration adjusted to correct for IP variations, then to calculate the RV of the star we simply need to find a way of calculating the RV from all the lines in the spectrum, and of averaging them in an optimal way (in the statistical sense). While one could in principle model each line independently, extract the wavelength corresponding to the center of each line from the model and calculate its RV relative to the theoretical wavelength, since there are 3000--4000 sharp lines per spectrum, this would be a computationally-heavy procedure. Instead, one condenses the information from all stellar lines present in the spectrum in an average stellar line, which is representative of the star. This is done by calculating the \textit{Cross-Correlation function} (CCF) between the spectra and a line list containing all lines selected for the RV calculation \citep{1996A&AS..119..373B}. In practice, the cross-correlation function is the convolution between the recorded spectra and a binary mask containing the wavelengths of the lines of interest, performed in the RV space (i.e., after shifting the mask over a range of RV). The binary mask can be upgraded to a mask containing the depth of each line so that the contribution of the depth on the precision is considered in the optimal construction of the average line \citep{2002A&A...388..632P}; when doing so we are assuming that all lines have very similar $FWHM$. The resulting average line is also called \textit{CCF} in what is an obvious abuse on nomenclature. For a slowly-rotating G- or K-type star (as are often the stars considered for high-precision RV searches), the CCF has a Gaussian shape, and the fit of a Gaussian function can efficiently deliver the center of the line in RV. At this point it is important to remember that a mismatch between the stellar line and the fitting function will not introduce an error on the value of the measured center, \textit{as long as the spectral line shape remains the same, i.e., as long as the IP does not change with time}. Any mismatch or systematic error introduced by the line shape in the calculation of the RV will be present in exactly the same way on every RV measurement, and will not impact our study of the RV variation, and of the RV \textit{precision}.

As one moves to later spectral types, i.e., to M dwarfs, the spectra become overpopulated with lines to the point the average distance between lines becomes smaller than the resolution of the spectrograph. The lines become \textit{blended} and the recorded spectra show a dense forest of overlapping lines, often creating regions of strong absorption and even a pseudo-continuum \citep[see, e.g.,][]{2016A&A...586A.101F}. An analysis of the spectra will still reveal that the information is in the slope of the spectra, naturally; however, cross-matching it with a mask will create an average line polluted by blends, with deep wings; more importantly, the procedure will not deliver the best precision. For M stars it is preferable to cross-correlate the recorded spectrum with a template derived either from a theoretical model or from an average spectrum calculated iteratively from the observations \citep[e.g.,][]{2015A&A...575A.119A, 2016Natur.536..437A}.

\subsubsection{IP modeling}

The IP control is a conceptually straightforward approach, but with its requirements on light injection stability and environmental control, it is impossible to apply on a general-purpose spectrograph, which is seldom built with these constraints in mind. The impossibility of implementation on typical slit spectrographs motivated the development of an alternative approach: to use a spectrograph that allows the IP to vary, but to devise observations and data analysis so that one can characterize the IP variation and correct for it. For this one needs a wavelength calibration device called a \textit{gas-cell}. A gas-cell is a container with a gas species (or group of species) with a well-characterized, high-resolution absorption spectrum, and that can be mechanically inserted before the slit of the spectrograph. This cell will then superimpose a wavelength reference on the stellar spectrum, and the product of the two will be registered by the spectrograph. The objective is to use the gas-cell spectrum to define the wavelength scale on top of the science spectrum, defining a wavelength calibration for each observation. This is possible because our gas-cell spectrum will be subject to the IP variations, and by comparing our gas-cell observations with the cell high-resolution wavelength spectrum, one can fully characterize the IP. One can then \textit{deconvolve} the IP from the measured spectrum to recover the stellar spectrum and measure its position relative to the reference. For a detailed description of the procedure the reader is referred to \cite{1996PASP..108..500B}.

The equation that represents the observed spectrum $A(\lambda)$ is
\begin{equation}\label{Eq:GasCell}
 A(\lambda) = \int [I_2(\lambda')\,S(\lambda + \delta \lambda)]\,IP(\lambda-\lambda')\,d\lambda' \, ,
\end{equation}
where $A(\lambda)$ is the relative intensity of the final spectrum as measured by the spectrograph, $I_2(\lambda)$ is the iodine cell spectrum, $S(\lambda)$ is the source spectrum and $IP(\lambda)$ is the instrumental profile, all as a function of wavelength $\lambda$. $\delta \lambda$ is the relative wavelength shift between the science and reference spectra. This equation represents how the observed spectrum is the product of the scientific/source spectrum by the reference spectrum, convolved with the spectrograph's instrumental profile. Our final scientific objective is to determine $\delta \lambda$. However, with the exception of $I_2(\lambda)$, all elements on the right-hand side of Eq.~(\ref{Eq:GasCell}) are unknown. As such, we will have to devise a clever observational scheme to determine each one of these unknowns, or spectra. The most common used recipe is as follows:
\begin{enumerate}
\setlength\itemsep{0.4em}	
 \item Measure the $I_2(\lambda)$ with a \textit{Fourier Transform Spectrograph} (FTS) to obtain a spectrum with a much higher resolution than that recorded with our spectrograph (usually $R=500{,}000$ or larger).
 \item Observe a line-less emission spectrum (e.g., lamp or bright hot star) with the spectrograph+cell and deconvolve the $I_2(\lambda)$ to obtain the $IP(\lambda)$.
 \item Observe the science target with a very high $SNR$ and \textit{without the $I_2$ cell}, to deconvolve the $IP(\lambda)$ from these observations and get $S(\lambda)$.
 \item Observe the science target \textit{with the $I_2$ cell}, and recover $\delta \lambda$ from the evaluation of Eq.~(\ref{Eq:GasCell}).
\end{enumerate}

Deriving RV through this method is clearly a complex process. It is subject to the fidelity with which each of the intermediate data products is obtained or determined. In particular the $IP$ reconstruction is a very delicate process; a careful parameterization should characterize the $IP$ as both a function of time and $\lambda$, with particular attention on how it depends on the position of the spectrum on the detector. Labour-intensive as it is, this methodology has been widely used to transform general-purpose spectrographs into efficient planet-hunting machines.

We can summarize the previous two sections in the following way:
\begin{itemize}
\setlength\itemsep{0.4em}	
 \item The \textbf{IP control} technique requires a stable spectrograph, both in light injection and thermo-mechanical stability. One can get the best of it by using a second channel to simultaneously record the spectrum of a reference calibration. \textbf{It minimizes the presence of instrumental RV shifts}.
 \item The \textbf{IP modeling technique} can be used on a general-purpose slit spectrograph, and \textbf{models and subtracts the IP variations that induce an RV shift}. It requires several on-sky calibrations and as such it is observationally expensive. 
\end{itemize}

When we are talking of the RV precision required to detect planets, of m/s (or no larger than 10 times that), we have to remember that these correspond to shifts of spectral lines at 1/1000 of the pixel size. This type of precision is incredible, and both techniques are undoubtedly successful by reaching this mark. Also, before comparing the two techniques it is important to note that many instruments can only use one of the methodologies due to the practical requirements they impose on the instrumentation. 

It is impossible to state which technique is capable of delivering the most precise RV without resorting to observational data. One technique minimizes the RV shifts without characterizing them, while the other characterizes them but through a complex process that has its own practical limitations. Probably the only way to settle this argument is to look at the best precision achieved on the two instruments that best embody the two techniques described here: HARPS \citep{2003Msngr.114...20M} and HIRES \citep{1994SPIE.2198..362V}. While HARPS reached a precision of 1\,m/s and better, HIRES floored at 2--3\,m/s of precision. These results were the subject of hot debates for a long time, but it is now solidly established the IP control technique is the only one able to deliver sub-m/s precision. 

So far we have been debating how to achieve the best precision on very general terms, but we have not mentioned that our own Earth is traveling in space, and the projection of our own RV along the line of sight of our stellar observations will shift the recorded spectra of the target. In order to correct the measured RV for this effect, one has to use the ephemerides of our own solar system to calculate the position of the Earth with great accuracy \citep[e.g.,][]{1988A&A...202..309B}.

\section{Current and future planet-hunting machines}

The great motivation for the development of precise RV spectrographs was the detection of extrasolar planets. Today the most precise planet-hunting machines are the spectrograph HARPS \citep{2003Msngr.114...20M}, installed at the 3.6-m telescope located at La Silla, and its northern twin HARPS-N \citep{2012SPIE.8446E..1VC}, installed at the TNG telescope at La Palma. These two spectrographs were developed to minimize the IP variation at an extreme level, and today yield a precision of 50--60\,cm/s. The main dispersive element is an echelle grating R4, with 31.6 gr/mm, operating at a blaze angle of $75^{\circ}$, and the cross dispersing is done with a grism. The spectrographs are fed by octogonal fibers, which have shown to have improved scrambling properties over the typical circular ones.

Until recently, the simultaneous reference and wavelength calibration on HARPS and HARPS-N was provided by a ThAr emission lamp. However, these are far from being ideal calibrators. The large dynamics and very different spatial density of the lines led to very different wavelength calibration stability as a function of wavelength, which translated ultimately in a different RV precision as a function of wavelength. Moreover, with the increased rarity of these lamps, there has been an active search for wavelength reference alternatives. 

The caracteristics of the perfect wavelength reference are very clear: 
\begin{itemize}
 \item should cover the whole spectral range of the spectrograph;
 \item the lines used in the calibration should have a $FWHM$ smaller than the spectrograph's resolution;
 \item the source should provide a high density of lines, up to one per 2--3 times the resolution element, and at a constant spacing;
 \item the wavelength of the lines should be precisely known and stable;
 \item the line intensities should be homegeneous, being close to the saturation but with a high dynamic range.
\end{itemize}

It is easy to conclude tha neither the ThAr lamp nor the I$_2$ cell get even close to these specifications, and nature cares not to provide such a level of homogeneity and fine-tuning in the form of atomic or molecular transitions. As such, the most recent advances focus on the development of back-illuminated Fabry--Perot cavities \citep[e.g.,][]{2011SPIE.8151E..47W, 2014PASP..126..445H, 2014A&A...569A..77R}, or of laser-frequency combs, in which a femtosecond laser is stabilized with an atomic clock to produce a series of modes that are subsequently filtered by a Fabry--Perot cavity \citep[e.g.,][]{2012SPIE.8446E..1WL}.  

With the current instrumentation, remarkable discoveries were made, like the Earth-mass planet in the habitable zone around our neighbor Proxima Centauri \citep{2016Natur.536..437A}. The rocky Earth-mass planets around Kepler-78 \citep{2013Natur.503..377P} and HD~219134 \citep{2015A&A...584A..72M} show how precise RV can be used in transiting planets to help determine a planet's bulk composition. Very importantly, the sub-m/s RV precision allowed us to start to uncover and characterize the population of Earth-mass planets, and we have already located stars that contain systems of exoplanets with up to 7 planets, like HD~10800 \citep[e.g.,][]{2011A&A...528A.112L}.

When several planets are present around a star and with orbital semi-amplitudes at the level of instrumental precision, it is very difficult to characterize their orbits, due to the large number of parameters to fit. This leads to the need of a large number of RV points per system, and motivated the development of more precise instruments. The forthcoming planet-hunter ESPRESSO \citep{2010SPIE.7735E..14P, 2014AN....335....8P} spearheads this quest, and will be able to reach an intrinsic RV precision of 10\,cm/s; along with the improved collecting capability of the VLT and associated photon noise contribution, ESPRESSO will be able to detect an Earth-mass planet inside the habitable zone around a solar-type star. This corresponds to a significant jump relative to the current 50\,cm/s precision that allows us to detect Earth-mass planets in orbits of only a couple of days, at most.

A different way of looking at the the challenge of detecting Earth-mass planets inside the habitable zone of their host stars is to turn to M dwarfs. The lightest stars experience a reflex motion $~$3 times larger than their GK companions, and their lower energy output draws the habitable zone roughly 3 times closer. This means that in order to detect an Earth-mass planet orbiting inside the habitable zone around an M dwarf, one needs only a precision of roughly 1\,m/s. However, since M dwarfs are faint in optical wavelengths, this scientific objective spurred the development of near-infrared instruments like SPIRou \citep{2014SPIE.9147E..15A}, NIRPS \citep{2016arXiv160801124C}, and CARMENES \citep{2014SPIE.9147E..1FQ}, just to cite a few among the many spectrographs of this category. With these spectrographs we will be able to extend our studies of planetary frequency from the first 100 M dwarfs surveyed with HARPS to a more complete sample in terms of stars and precision. Last but not least, in a more distant future, the European Extremely Large Telescope (E-ELT) will have at least an instrument --- HIRES --- capable of delivering precise RV at a 10\,cm/s precision or better, and will enable RV studies on stars that are otherwise discarded due to photon noise limitations.  

This contributions tries to cover the main aspects behind RV precision and its association to planetary studies. However, with the number of developments in the latest couple of decades, with 3 hours of lectures we could only skim the surface of all the topics that are there to discuss. And after the derivation of precise RVs comes its interpretation, for which the points discussed here will be of great use. But that's a story for another time.

\begin{acknowledgement}
I acknowledge support by Funda\c{c}\~ao para a Ci\^encia e a Tecnologia (FCT) through Investigador FCT contract of reference IF/01037/2013/CP1191/CT0001, and POPH/FSE (EC) by FEDER funding through the program ``Programa Operacional de Factores de Competitividade - COMPETE''. I further acknowledge support from FCT in the form of an exploratory project of reference also IF/01037/2013/CP1191/CT0001. I acknowledge \textit{Peter Sport Caf\'{e}} for showing me what a good gin tonic tastes like, and M\'{a}rio Jo\~{a}o Monteiro, Mahmoudreza Oshagh and Vardan Adibekyan for reminding me that good science requires good statistics.  
\end{acknowledgement}

\bibliographystyle{apj}
\bibliography{Mybibliog}

\end{document}